\documentclass[prb,preprint]{revtex4-1} 

\usepackage{latexsym}
\usepackage{amsmath}
\usepackage{verbatim}

\newcommand{\beq}{\begin{quote}}

\newcommand{\enq}{\end{quote}}
\newcommand{\be}{\begin{equation}}
\newcommand{\en}{\end{equation}}

\begin{document}

\title{Comment  on  QBism and locality in quantum mechanics}
\author{Michael Nauenberg}

\affiliation{Physics Dept. University of California, Santa Cruz, CA 95064}

\maketitle

Felix Bloch recounted  that after Erwin Schr\"{o}dinger introduced his  wave function $\psi$,  
a verse circulated among his fellow students:
\begin{quote}
Erwin with his psi can do\\
Calculations quite a few.\\
But one thing has not been seen:\\
Just what does $\psi$ really mean? \cite{bloch}
\end{quote} 

According to Fuchs {\it et al.},\cite{mermin} the question regarding the meaning of $\psi$, which was raised shortly after the formulation of quantum mechanics, has remained unsolved.  Originally, Schr\"{o}dinger  had  proposed that $|\psi(x,t)|^2$ represented the charge density  of the electron at  time $t$ in an interval between $x$ and $x+dx$, but he soon realized that this interpretation  ran into difficulties even for a free electron, because his equation for $\psi$ implied that $\psi$ would   spread as a function of time.\cite{sch}  But experimentally it was well known that the electron remains localized like a point particle. Shortly afterwards,  Max Born introduced the interpretation that  $|\psi (x,t) |^2$ is  the {\it probability} density  for an electron to be found at time $t$ in this interval.\cite{born, born2}  In his own words, ``the motion of particles follows  the laws of probability,  but the probability itself spreads in harmony with causal laws,'' and in a footnote he clarified his statement with the remark that ``the knowledge of a state at all points in one moment, determines the state at all times.''\cite{born2}  By probability, it  is important  to emphasize here that Born meant  the {\it frequency} of different  outcomes predicted by $|\psi|^2$,  after a given experiment is repeated multiple times under identical  initial conditions.  These  conditions,  and the various possible final outcomes are experimentally established by measurement devices that can permanently record such events by a macroscopic and time irreversible process.  Virtually all experiments in quantum mechanics have these features, whether the measuring apparatus consists of an ancient Geiger counter or a modern detector.  The observer's main role  is to design and build the devices required for a given experiment,  to calculate the frequency or  probability for all possible outcomes according to quantum mechanics, encapsulated in $\psi$, and to publish the results.  Up to date,  experiments in the micro-world  have always confirmed Born's frequency interpretation of $|\psi|^2$.

By taking a  {\it subjective} or Bayesian view of probabiltiy,  the QBist interpretation of quantum mechanics, described in the article by  Fuchs {\it et al.}, effectively denies that the outcome of experiments are described by permanent records, independently of the views of any particular  observer or so-called ``agent.''  Although Fuchs {\it et al.\ }agree that quantum states determine probabilities through the Born rule, they assert  without any justification that,  ``since probabilities are the personal judgements of an agent, it follows that a quantum assignment is also a personal judgment of the agent assigning that state'' (p.~749).  But for any  experiment  these  agents  calculate the same values for $\psi$, and therefore they all  obtain the same probability $|\psi |^2$  to observe the possible outcomes of their experiment.  In their article, Fuchs {\it et al.\ }do not provide a single experiment that falsifies  this conventional view of quantum mechanics, proposing, instead, their QBism interpretation of quantum mechanics without providing a single experiment that validates it. 

For an example, consider the eponymous double-slit experiment discussed in all elementary textbooks on quantum mechanics.  At sufficiently low intensity, a light beam containing only a few photons impinging on the slit with a photographic screen behind it records the individual impacts of these photon.  At  first these photons appear randomly scattered on this screen, but after a large number  of them are recorded, a pattern forms corresponding to the well known interference pattern that forms on the screen when a high intensity light beam is  transmitted through the slits.  It has been demonstrated in numerous experiments that this interference pattern corresponds precisely to the frequency or probability distribution evaluated according to $|\psi|^2$  that individual photons land on a given spot on the screen.

Regarding the question addressed by Fuchs {\it et al.\ }on whether quantum mechanics is nonlocal, consider the correlation between the spin states of two electrons with total spin angular momentum zero.  This is the main spin component 
in the ground state of the helium atom, and there has never been  any issue about locality concerning this correlation,  because the two electrons are confined spatially to the domain of the atom. Now suppose that these two electrons are ionized simultaneously without affecting their total spin state, and the two electron move apart. Then quantum mechanics predicts that in the absence of any new interaction or entanglement with other particles (e.g., the environment) these  correlations remain the same, even after these electrons are separated by a large distance.  What would  be ``spooky,'' using Einstein's terminology, is that the initial two-electron spin correlation would change under these conditions.  Hence, contrary to the claim of Fuch's {\it et al.\ }(p. 751), quantum mechanics  does  assign correlations to space-like separated events. Unlike in classical mechanics, however,  the observed spin state of an  electron depends also on the measuring device, which can be altered during the time that these electrons travel to  reach these devices in a correlation
experiment, leading, from the viewpoint of  {\it reality} in classical physics,  to an  {\it apparent} non-locality. Correlated events can be recorded by detectors at space-like separations,and afterwards sent to a single ÒagentÓ, as it is readily done in
practice. Hence, the question of locality is not resolved by fiat as  claimed by Fuchs et al. in their QBist interpretation of quantum mechanics.

Fuchs {\it et al.\ }conclude that: ``\dots quantum mechanics itself does not deal directly with the objective world; it deals with the experiences of that objective world that belong to whatever particular agent is making use of the quantum theory'' (p.~750).  But in his lengthy correspondence with Einstein, Born already had emphasized that in practice, classical mechanics also is a statistical theory, because the initial conditions and the final outcome are never know with absolute precision.\cite{born3}  In particular, in systems obeying chaotic dynamics, sensitivity to initial conditions implies that the outcome can be completely random.  The essential difference in quantum mechanics, however, is that  the precision of initial conditions is limited by Heisenberg's uncertainty principle $\Delta p \Delta x \geq \hbar/2$.   Hence, contrary to Fuchs {\it et al}, quantum theory deals with the objective world as  directly as does classical mechanics.

\end{document}